\documentclass{eas}
\usepackage{graphicx}
\begin{document}
\title{CEMP-\textsc{s} and CEMP-\textsc{s/r} stars: last update}
\author{Bisterzo, S.,}\address{Department of Physics, University of Turin, Italy}
\secondaddress{INAF - Astrophysical Observatory Turin, Turin, Italy}
\author{Gallino, R.,}\sameaddress{1}
\secondaddress{B2FH Association-c/o Strada Osservatorio 20, 10023 Turin, Italy}
\author{Straniero, O.,}\address{INAF - Osservatorio Astronomico di Collurania, Teramo, Italy}
\author{Cristallo, S.,}\sameaddress{3}
\author{K{\"a}ppeler, F.,}\address{KIT - Karlsruhe Institute of Technology, Karlsruhe, Germany}
\author{Wiescher, M.,}\address{Department of Physics, University of Notre Dame, Notre Dame, IN}

\begin{abstract}

We provide an updated discussion of the sample of CEMP-s and CEMP-s/r stars collected 
from the literature. Observations are compared with the theoretical 
nucleosynthesis models of asymptotic giant branch (AGB) stars presented 
by Bisterzo {\em et al.\/} (\cite{bisterzo10,bisterzo11,bisterzo12}), in the light
of the most recent spectroscopic results. 

\end{abstract}
\maketitle
\section{Introduction and open problems}

Metal-poor stars are key objects to investigate the origin of the elements in the early
Universe. 
The very high resolution and signal-to-noise ratio reached in the last decades by different survey 
projects (HET/HRS, Keck/HIRES, Magellan/MIKE, Subaru/HDS, and VLT/UVES) allowed the 
discovery of several chemically peculiar objects. 
\\
Known metal-poor stars are classified in several groups, following their chemical
peculiarities (Beers {\em et al.\/} \cite{beers05}; Aoki {\em et al.\/} \cite{aoki07}).

About 10\% of halo stars show an enhanced carbon abundance (Carbon Enhanced Metal-Poor, 
CEMP); this fraction increases up to $\sim$30\% by decreasing metallicity ([Fe/H] $<$ 
$-$3). For metallicities lower than [Fe/H] $\sim$ $-$4.5, the analysis is still statistically 
insignificant with only four objects known (three of which are C-rich; see Yong {\em et al.\/} 
\cite{yong13} and references therein).
CEMP stars are distinguished in four subclasses based on the degree of enhancement detected
among neutron capture elements: CEMP-s (s-process enriched; [Ba/Fe] $>$ 1 and [Ba/Eu] $>$ 0.5), 
CEMP-s/r (s- and r-process enriched; [Ba/Fe] $>$ 1 and 0 $<$ [Ba/Eu] $<$ 0.5), CEMP-r 
(r-process enriched; [Eu/Fe] $>$ 0.3 and [Ba/Eu] $<$ 0), CEMP-no (with sub-solar s-process abundances; 
[Eu,Ba/Fe] $<$ 0), and CEMP-$\alpha$ stars (with a large excess of C, N, O and $\alpha$-elements).
Note that most of the metal-poor r-rich stars known are not carbon enhanced and are  
classified as r-II (with [Eu/Fe] $\ge$ 1) or r-I (0.3 $\le$ [Eu/Fe] $\le$ 1).
About 80\% of CEMP stars are CEMP-s, and about half of CEMP-s with measured Eu are also r-rich 
(CEMP-s/r). 
We focus our analysis on these two last classes of stars.
\\
The s-process mainly occurs in stars of low initial mass ($M$ $\sim$ 1 -- 3 $M_\odot$) 
during their thermally pulsing asymptotic giant branch 
phase (TP-AGB), at rather low neutron densities ($n_n$ $\sim$10$^7$ neutrons/cm$^3$). 
The major neutron source of the s-process in AGBs is the $^{13}$C($\alpha$, n)$^{16}$O reaction, 
which burns radiatively (at T $\sim$ 0.9$\times$10$^8$ K) in a thin layer at the top of the 
He-intershell (called $^{13}$C-pocket) during the interpulse periods. A second neutron source is marginally 
activated during the convective thermal pulses (T $>$ 2.5$\times$10$^8$ K), mainly affecting 
isotopes close to the branching points.
An additional s-process component comes from
massive stars during core He and shell C burning ($n_n$ $\sim$10$^{12}$ neutrons/cm$^3$) and 
contributes to neutron capture isotopes up to A $\sim$ 90. The r-process requires high neutron 
densities (up to $n_n$ $\sim$10$^{22}$ neutrons/cm$^3$) to produce neutron rich isotopes far from 
the stability valley and its origin is currently attributed to explosive nucleosynthesis in
massive stars. 
For extended reviews on the s- and r-processes we refer to K{\"a}ppeler {\em et al.\/} 
(\cite{kaeppeler11}) and Thielemann {\em et al.\/} (\cite{thielemann11}).

The carbon enhancement coupled with high s-process abundances detected in CEMP-s is 
associated to binary systems, where the more massive companion (now an invisible white dwarf) 
evolved faster through the TP-AGB and polluted the observed CEMP-s star 
 by mass transfer through efficient stellar winds. Indeed CEMP-s
are old main-sequence or giant stars, with low metallicity ([Fe/H] $\le$ $−$2) and low initial 
mass ($M$ $<$ 0.9 $M_\odot$), which lie far from the TP-AGB phase: as sustained by radial 
velocity studies by Lucatello {\em et al.\/} (\cite{lucatello05}), their
surface composition has to be modified by accretion of material from one (or more) companion(s).
\\
 Large uncertainties still affect AGB models and s-process 
nucleosynthesis, as the evaluation of the mass-loss, the efficiency of the third dredge-up (TDU, 
a mixing episode that permits partial mixing processes between material of the He-intershell and 
the convective envelope), the formation of the $^{13}$C-pocket (Herwig \cite{herwig05}; Straniero,
Gallino \& Cristallo \cite{straniero06}). In case of binary systems, the amount of mass
accreted and the distance between the two stars also influence the s-enhancement. 
In this context, we study the behavior of the ratios between the three s-process peaks, ls (light-s
elements at N = 50, i.e. Sr-Y-Zr), hs (heavy-s elements at N = 82, i.e. Ba-La-Ce) and Pb (at N = 126):
the [hs/ls] and [Pb/hs] ratios are extremely valuable indexes for the s-process as they are 
independent of both the TDU efficiency in the AGB star and the dilution of the AGB material on 
the companion. 
\\
In addition, the C and s-material observed on 
the surface is affected by internal mixing occurring during the stellar life: e.g., radiative 
acceleration competes with gravitational settling during the lengthy main-sequence phase, 
while thermohaline instabilities may reach deep layers on a shorter time-scale 
if not prevented by the other two processes.
The resulting abundance alteration is very difficult to estimate (e.g., Richard {\em et al.\/} 
\cite{richard02}; Vauclair \cite{vauclair04}; Stancliffe {\em et al.\/} \cite{stancliffe07}; 
Stancliffe \& Glebbeek \cite{stancliffe08}; Vauclair \& Th{\'e}ado \cite{vauclair12}), especially 
if rotation or magnetic fields are included in the analysis.
The comparison between theoretical predictions and observations helps to establish the efficiency 
of non-convective mixing in the envelope of the observed star during their main-sequence 
phase. Present results by Bisterzo {\em et al.\/} (\cite{bisterzo08,bisterzo11}) seem to indicate 
that no efficient mixing takes place in main-sequence stars. This agrees with model calculations 
by Thompson {\em et al.\/} (\cite{thompson08}), who showed that gravitational settling
can confine the efficiency of thermohaline mixing in these low-mass
metal-poor stars. However, the statistic was limited to seventeen main-sequence stars. 
For stars on the red giant branch, having undergone the first
dredge-up episode (FDU), all mixing processes occurred during
the main-sequence phase are erased.
\\
As already found for disk stars (e.g., post-AGB, Ba stars; Busso {\em et al.\/} \cite{busso01}; Abia
{\em et al.\/} \cite{abia02}), for a given
metallicity a range of $^{13}$C-pocket strengths has been hypothesized to interpret the observations 
in CEMP-s stars. A clear definition of the properties of the mixing processes at radiative/convective 
interfaces that lead to the formation of the $^{13}$C-pocket has not been reached yet. Moreover, models 
including rotation, gravity waves or magnetic fields may influence the formation 
of the $^{13}$C-pocket (Langer {\em et al.\/} \cite{langer99}; Denissenkov \& Tout \cite{denissenkov03}; 
Herwig, Langer \& Lugaro \cite{herwig03}; Siess, Goriely \& Langer \cite{siess04}; Piersanti {\em et al.\/} 
\cite{piersanti13}). This translates into different s-process distributions. 

CEMP-s/r are among the most enigmatic stars because the s- and r- processes are commonly related 
to different nucleosynthesis environments. 
Ba (or La) and Eu are adopted as reference elements to investigate the competition between 
the two processes, given that $\sim$80\% of solar Ba (or $\sim$70\% of solar La) is synthesized by 
AGBs, while $\sim$94\% of solar Eu is produced by the r-process. 
A pure s-process predicts [Ba,La/Eu] $\sim$ 1, while CEMP-s/r show an observed [Ba,La/Eu] ratio down
to zero. It is ascertained that a pure s-process is not sufficient to explain the observations in
CEMP-s/r and different scenarios have been proposed in literature (Cohen {\em et al.\/} \cite{cohen03} 
and references therein; Zijlstra \cite{zijlstra04};
Jonsell {\em et al.\/} \cite{jonsell06}; Aoki {\em et al.\/} \cite{aoki06};
Sneden, Cowan \& Gallino \cite{sneden08}; Bisterzo {\em et al.\/} 
\cite{bisterzo09}; Lugaro {\em et al.\/} \cite{lugaro12}).
The origin of CEMP-s/r is highly debated and an unanimous interpretation is not given by the 
scientific community.
\\
Our hypothesis has been discussed in detail by Bisterzo {\em et al.\/} (\cite{bisterzo11}). 
We assume that the r-enhancement detected in peculiar stars with very low metallicities may be 
due to local Supernova(e) explosion(s), leading to an r-enrichment of molecular clouds from which 
CEMP-s/r stars may have formed. The observed CEMP-s/r and the more massive AGB companion belong to 
the binary system and have the same initial r-enhancement. The more massive star is supposed to 
evolve through the TP-AGB phase, synthesizing s-elements and polluting the observed companion through
stellar winds. 
The choice of the initial r-enhancement (scaled to Eu) for elements heavier than Ba is based on
the solar isotopic r-process contributions deduced with the residual method. 
The hypothesis of different initial r-process enhancements derives from on the spread observed in 
[Eu/Fe] in the Galactic halo (see SAGA database; Suda {\em et al.\/} \cite{suda08,suda11}).
Despite a few spectroscopic 
binaries, r-rich objects do not belong to binary systems, suggesting that their r-process 
enhancement reveals the chemical inhomogeneity of the interstellar medium of the Galactic
halo (Hansen {\em et al.\/} \cite{hansen11}). Possibly, the strong r-enhancement seen in
r-II stars results from the pollution of the pristine molecular cloud by a neighbor Supernova
(Aoki {\em et al.\/} \cite{aoki10}). 
\\
While the origin of neutron capture elements lighter than Ba is complex, likely resulting from the 
competition of a multiplicity of r-process components (Travaglio {\em et al.\/} 
\cite{travaglio04}; Honda {\em et al.\/} \cite{honda06,honda07}), r-abundance distribution of elements 
heavier than Ba observed in r-rich stars exhibits a scaled-solar r-process pattern (Sneden {\em et al.\/} 
\cite{sneden08}). As shown by
Bisterzo {\em et al.\/} (\cite{bisterzo11}, and references therein), an initial r-enhancement based on the 
same solar-scaled r-abundances distribution plausibly plausibly explains the r-enrichment 
in CEMP-s/r stars within uncertainties.
This may imply that the r-elements detected in CEMP-s/r and in r-rich stars have a common origin. 
Seventeen CEMP-s/r stars with large r-enhancement ([Eu/Fe] $>$ 1.5 and 0 $<$ [Ba,La/Eu] 
$<$ 0.5; called CEMP-sII/rII) and five CEMP-s/r with mild r-enhancement (1.0 $<$ [Eu/Fe] $<$ 1.5 
and 0 $<$ [Ba,La/Eu] $<$ 0.5; called CEMP-sII/rI) were discussed by Bisterzo {\em et al.\/} 
(\cite{bisterzo11,bisterzo12}).
The current fraction of known r-II stars\footnote{CS 22892-052, 
CS 22183-031,
CS 22953-003, 
CS 29491-069, 
CS 29497-004, 
CS 31078-018, 
CS 31082-001, 
HE 1219-0312 by Hayek et al. (2009); 
HE 1523-0901, 
HE 2327-5642, 
SDSS J2357-0052, 
HE 2224+0143, 
HE 1127-1143,
HE 0432-0923, 
HE 0430-4901, 
CS 22875-029, 
CS 22888-047;  
see SAGA database for references.}
seems compatible with the number of CEMP-sII/rII stars.
However, the metallicity of r-II stars is on average $\sim$0.3 dex lower than that of CEMP-sII/rII
stars: r-II stars range from $-$3.4 $\le$ [Fe/H] $\le$ $-$2.5 ([Fe/H]$_{\rm averaged}$ = $-$2.8)
and CEMP-sII/rII from $-$3.0 $\le$ [Fe/H] $\le$ $-$2.0 ([Fe/H]$_{\rm averaged}$ = $-$2.5).
Aoki {\em et al.\/} (\cite{aoki10}) highlight that, while more observations are needed for 
[Fe/H] $\le$ $-$3, the upper metallicity limit given for r-II stars
is well constrained by the large sample of objects detected for [Fe/H] $\ge$ $-$2.5. 
The number of r-I stars is greater than r-II stars and covers a larger metallicity range (up 
to [Fe/H] $\sim$ $-$1.5; e.g., SAGA database and Aoki {\em et al.\/} \cite{aoki10}). 
Five r-I stars have been analyzed at very-high resolution\footnote{BD +173248,
CS 30306-132, 
HE0420+0123, 
HD 221170, 
HD 115444.}, and the abundance pattern of neutron-capture elements heavier than Ba is quite
similar to that of the solar r-process component.
Their lower Eu abundance is likely more difficult to be disentangled from the averaged 
enrichment seen in the halo ([Eu/Fe] $\sim$ 0.5), which results from a multiplicity of SNe events 
that exploded during the Galactic lifetime. 
\\
Masseron {\em et al.\/} (\cite{masseron10}) found a linear correlation between [Ba/Fe] and [Eu/Fe]
observed in CEMP-s/r, and Lugaro {\em et al.\/} (\cite{lugaro12}) suggested a sort of connection 
in the origin of Ba and Eu in these stars.
Current low-metallicity AGB models do not reach the physical conditions needed to
increase [Ba/Eu], even under extreme conditions (e.g., PIE episodes, see Cristallo {\em et al.\/} 
\cite{cristallo09} and references therein). 
Moreover, solar abundances and neutron capture rates close to La and Eu isotopes (corresponding 
to a theoretical [La/Eu] ratio for pure s-process material close to 1 dex) are known with high 
accuracy. 
Therefore, we underline that, at present, AGB predictions can not interpret [Ba/Eu] $\sim$ 0 
together with [Ba/Fe] $\sim$ 2, without the assumption of an initial r-enhancement.

In the next Section we examine the results found by Bisterzo {\em et al.\/} 
(\cite{bisterzo10,bisterzo11,bisterzo12}), in the light
of the most recent spectroscopic information.

\section{Newly discovered CEMP-s and CEMP-s/r stars, updated results and future prospects}

Since the last two years, new CEMP-s and CEMP-s/r stars have been discovered and previous 
spectroscopic analyses have been complemented with more detected elements. 
We discuss these updates in the light of Bisterzo {\em et al.\/} 
(\cite{bisterzo11,bisterzo12}) using the AGB models described by Bisterzo {\em et al.\/} 
(\cite{bisterzo10}).

\begin{figure}[h]
\includegraphics[angle=-90,width=14pc]{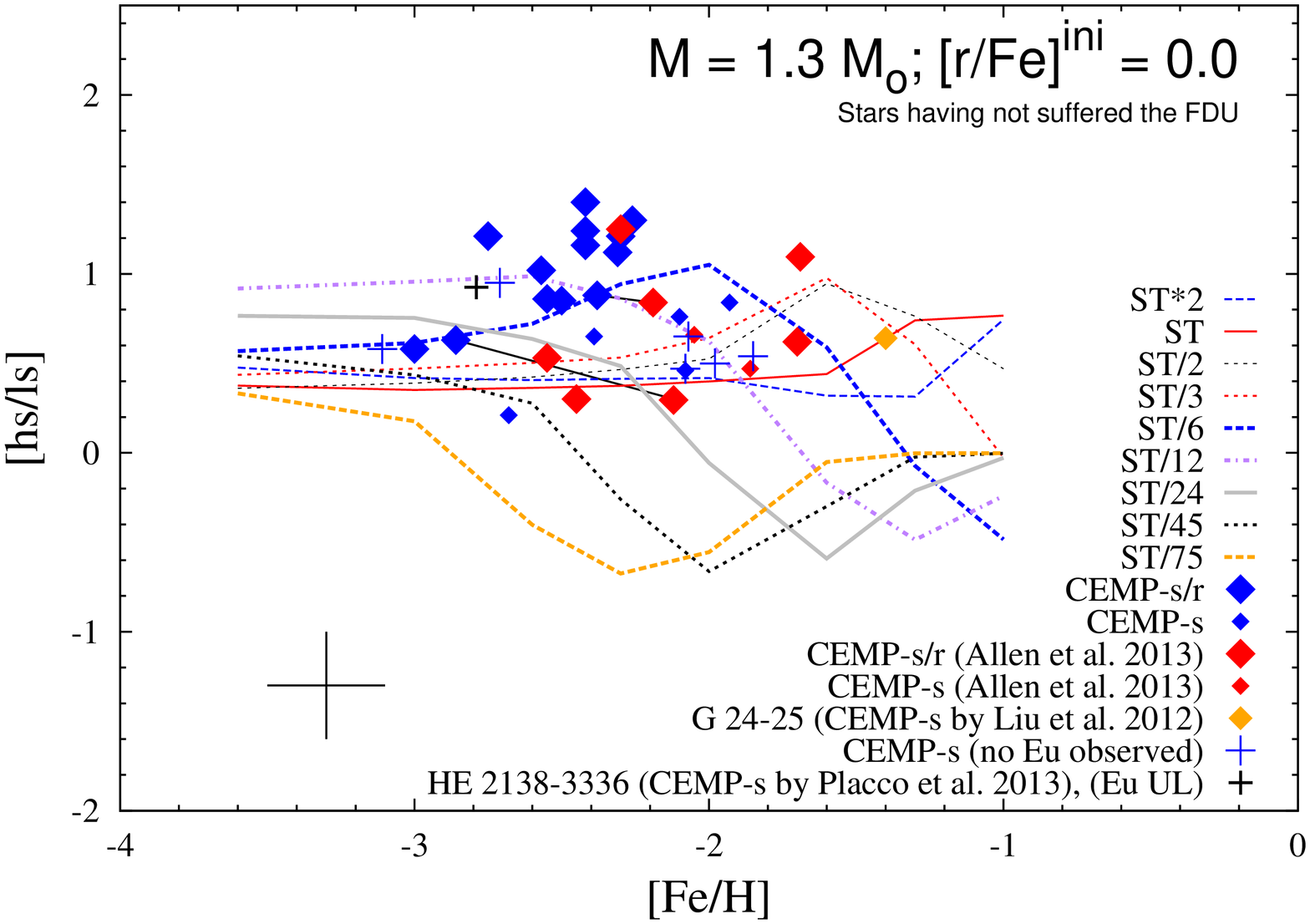}\hspace{2pc} 
\includegraphics[angle=-90,width=14pc]{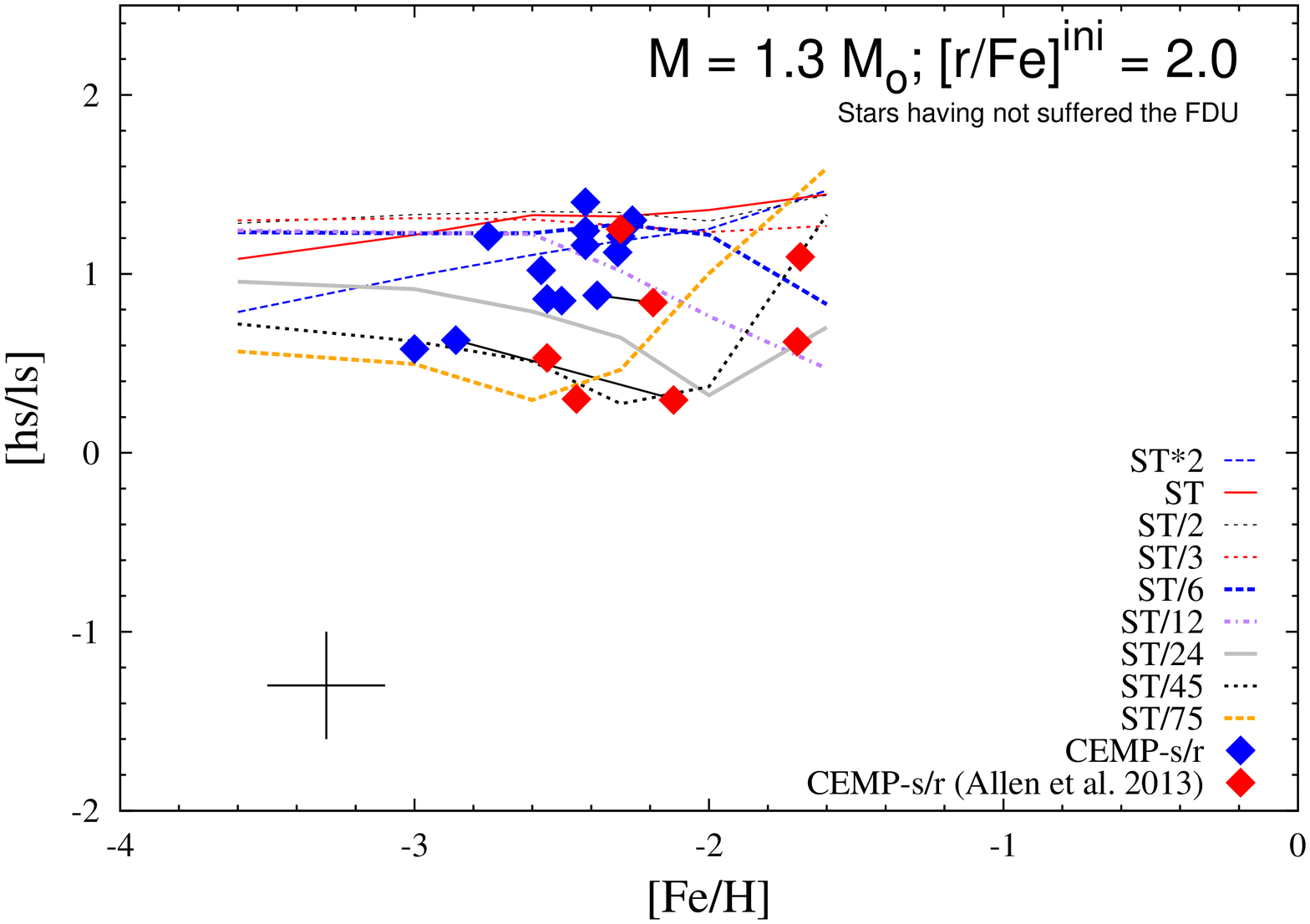}\hspace{2pc} 
\includegraphics[angle=-90,width=14pc]{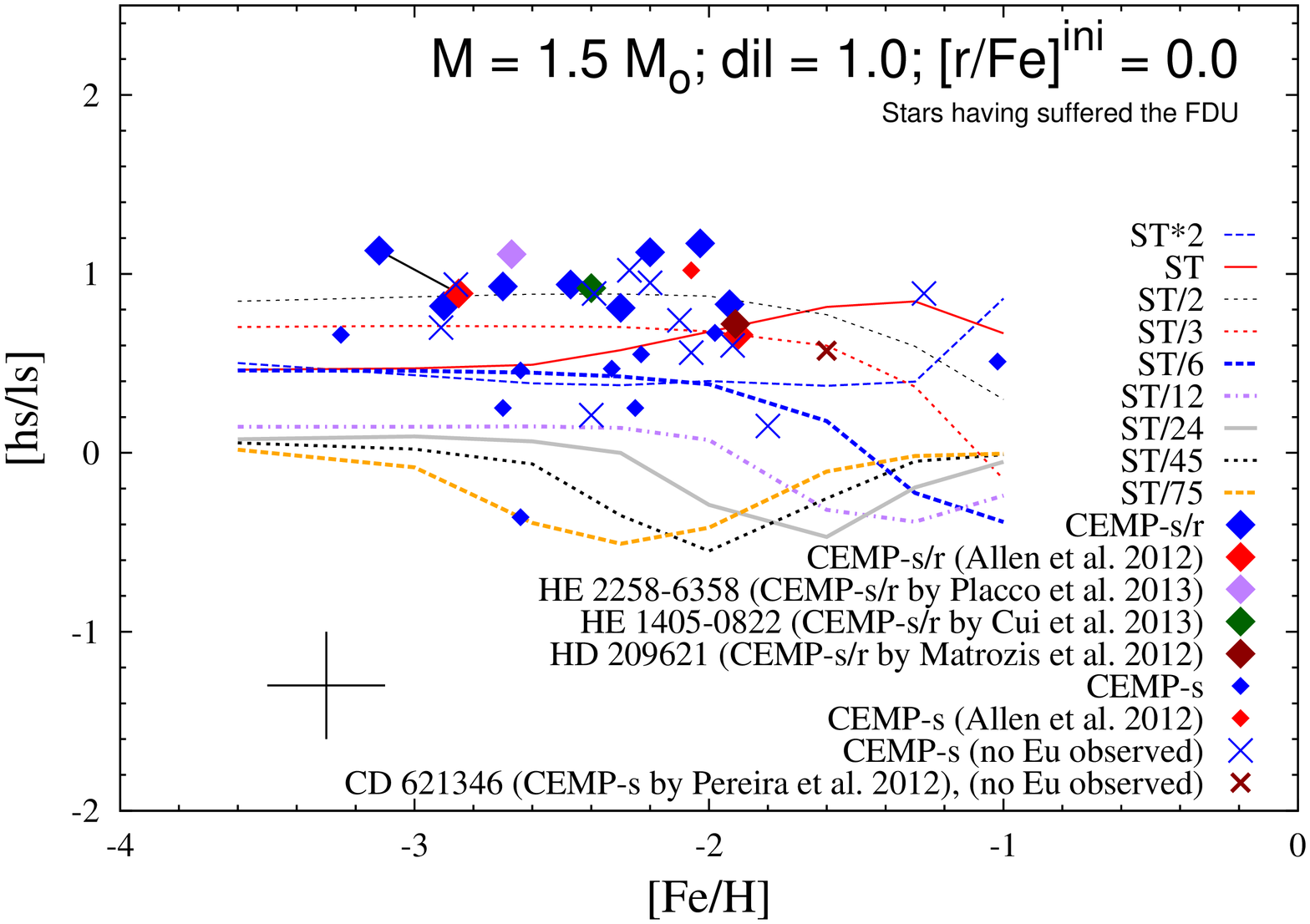}\hspace{2pc} 
\includegraphics[angle=-90,width=14pc]{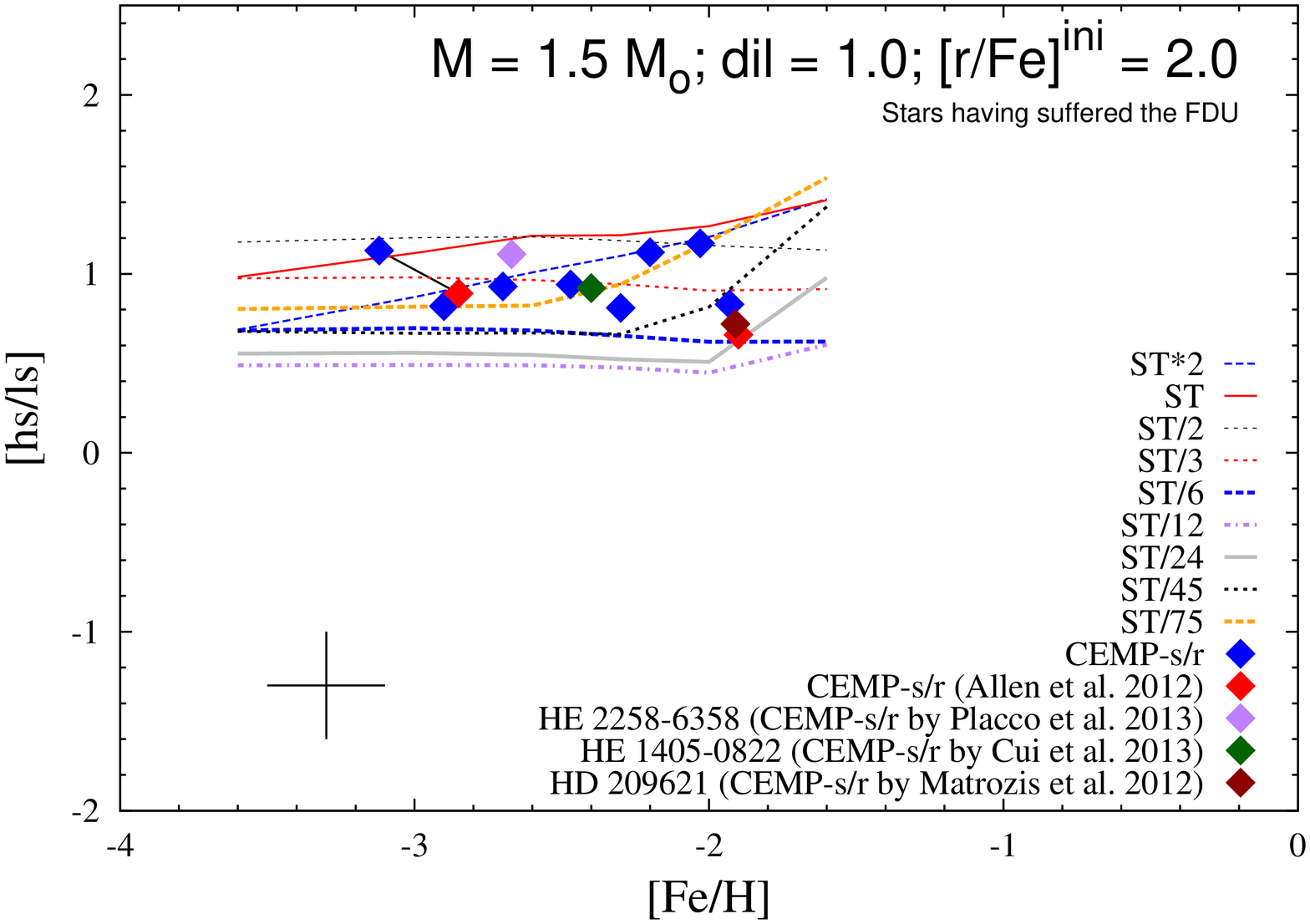}\hspace{2pc} 
\caption{\label{Fig1} {\bf Right panels}: 
theoretical predictions of {[hs/ls]} versus metallicity for
AGB models with initial mass $M$ = 1.3 and 1.5 $M_\odot$ (top and bottom panels,
respectively). A range of $^{13}$C-pockets (cases ST$\times$2 -- ST/45) 
is needed to interpret the observations of CEMP-s and CEMP-s/r stars. We display
with blue symbols the sample of stars analyzed by Bisterzo {\em et al.\/} 
(\cite{bisterzo11}). 
As made by Bisterzo {\em et al.\/} \cite{bisterzo11}, main-sequence/turn-off and 
subgiants not having suffered FDU are compared with 
$M$ = 1.3 $M_\odot$ model (top panel); subgiants/giants are 
compared with $M$ =  1.5 $M_\odot$ models and dil = 1.0 dex to simulate FDU mixing
(bottom panel).
CEMP-s stars without europium detection are indicated by blue plus symbols.
With respect to previous works, the following objects are added to the sample:
Allen {\em et al.\/} \cite{allen12} studied 9 CEMP-s/r stars (big red 
diamonds), and three CEMP-s stars (small red diamonds);
Placco {\em et al.\/} \cite{placco13} discovered a new CEMP-s/r giant (HE 2258--6358;
violet big diamond) and a CEMP-s main-sequence turnoff star (HE 2138--3336; black
plus symbol, for which a low upper limit for Eu is detected);
the new CEMP-s/r giant HE 1405--0822 by Cui {\em et al.\/} \cite{cui13};
the CH star G 24--25 by Liu {\em et al.\/} \cite{liu12};
the CEMP-s/r giant HD 209621 by Matrozis {\em et al.\/} \cite{matrozis12};
the CEMP-s star CD 621346 by Pereira {\em et al.\/} \cite{pereira12}.
{\bf Left panels}: same as the right panels, but CEMP-s/r are compared with AGB models
with high initial r-enhancement ({[r/Fe]}$^{\rm ini}$ = 2.0).
Typical error bars are {[hs/ls]} $\sim$ 0.3; {[Fe/H]} $\sim$ 0.2.}
\end{figure}

\begin{figure}[h]
\includegraphics[angle=-90,width=14pc]{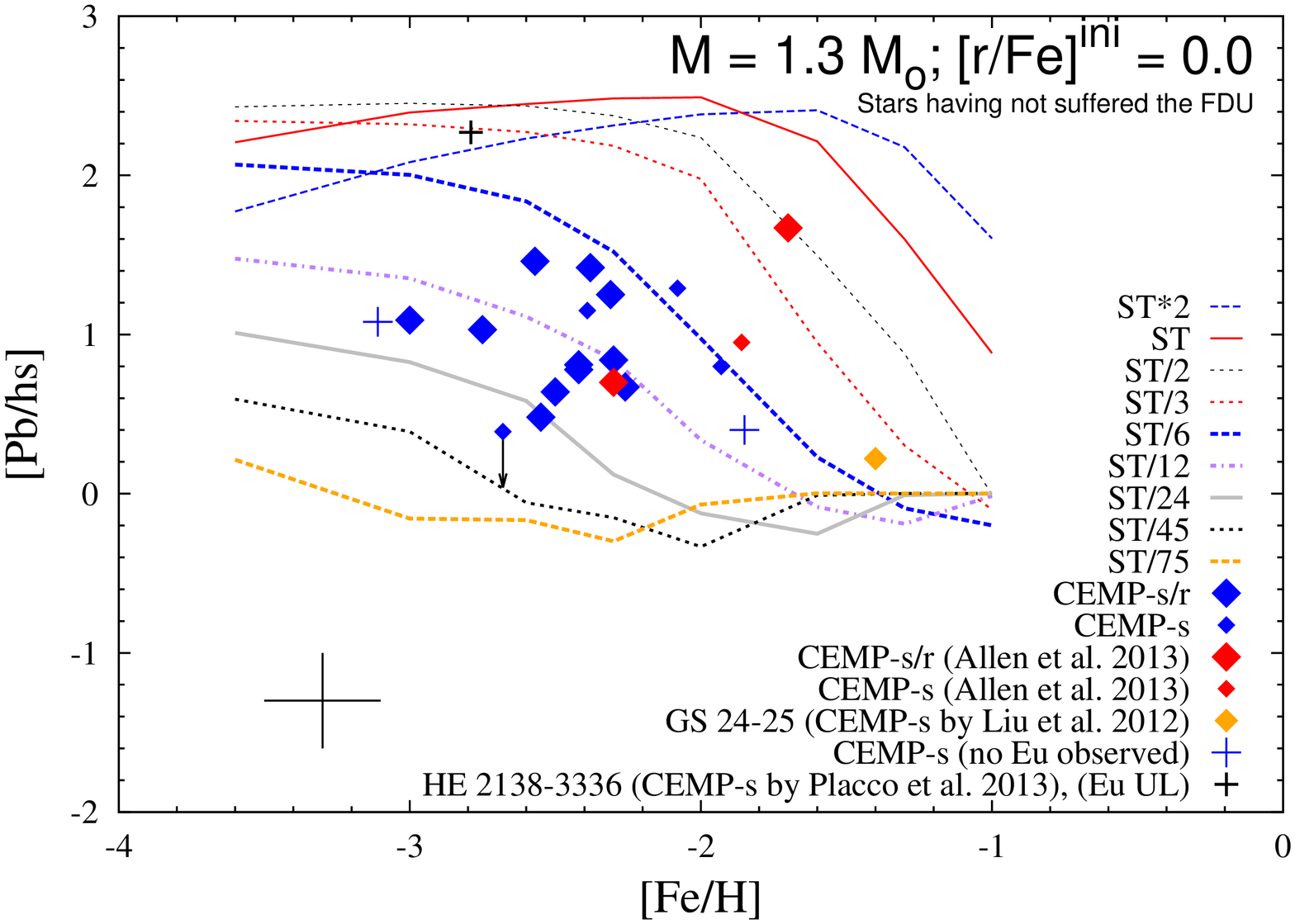}\hspace{2pc}
\includegraphics[angle=-90,width=14pc]{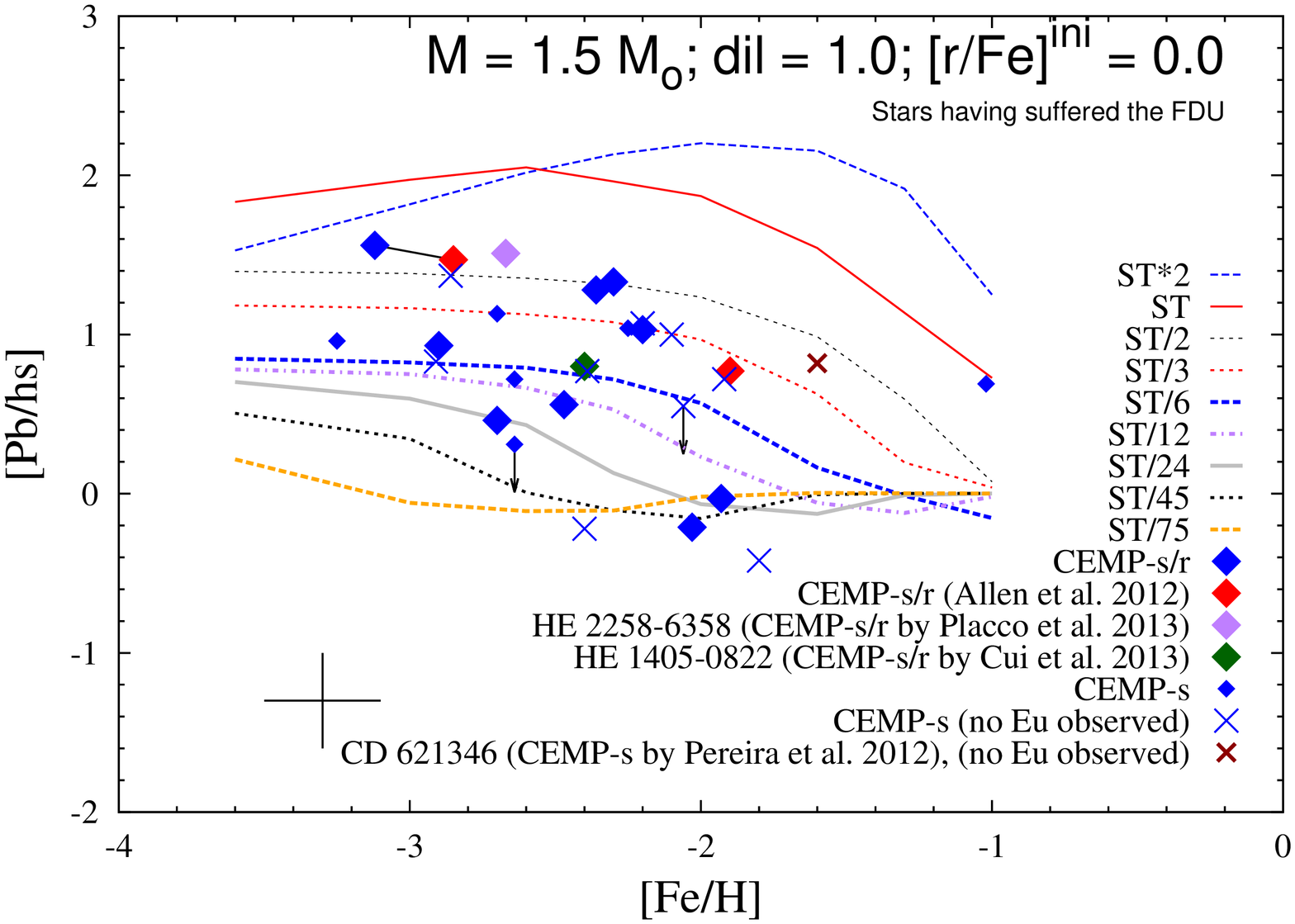}\hspace{2pc} 
\caption{\label{Fig2} The same as Fig.~\ref{Fig1}, right panels, but for [Pb/hs]
versus metallicity.}
\end{figure}

CEMP-s/r stars appear to have on average higher [hs/Fe] than CEMP-s stars.
This seems to agree with AGB model predictions if a high initial r-process enhancement of 
the molecular cloud is adopted ([r/Fe]$^{ini}$ = 2.0; see discussion by Bisterzo {\em et 
al.\/} \cite{bisterzo11}). In particular, the predicted [hs/ls] ratios reach values as high 
as 1.3 dex\footnote{An [r/Fe]$^{ini}$ = 2.0 may affect the final [hs/Fe], because $\sim$30\% 
of solar La, $\sim$40\% of solar Nd and $\sim$70\% of solar Sm are synthesized by the r-process.
Note that we exclude Sr from the ls-elements and Ba from the hs-peak because they are mainly
affected by higher spectroscopic uncertainties (Busso et al. 1995)
due to NLTE effects (Short \& Hauschildt \cite{sh06}; Mashonkina {\em et al.\/} \cite{mashonkina08}; 
Andrievsky  {\em et al.\/} \cite{andr11}), especially by decreasing the
metallicity.}. 
\\
In Fig.~\ref{Fig1} we show the behavior of [hs/ls] versus metallicity for AGB models 
with initial mass $M$ = 1.3 and 1.5 $M_\odot$ (top and bottom panels, respectively) 
and a range of $^{13}$C-pockets, compared with observations of CEMP-s and 
CEMP-s/r stars available in literature. 
In the right panels no initial r-enhancement is assumed, while in the left panels
CEMP-s/r stars are compared to AGB models with an initial r-enhancement of
[r/Fe]$^{\rm ini}$ = 2.0. 
We made an accurate choice of the CEMP-s and CEMP-s/r sample, based on the number of s-process
elements detected for each star (see Table 2 by Bisterzo {\em et al.\/} \cite{bisterzo11};
blue symbols). If only Sr and Ba are detected among the s-elements, the star is not displayed
in Fig.~\ref{Fig1}.
Fig.~\ref{Fig2} shows the same as the right panels of Fig.~\ref{Fig1}, but for [Pb/hs]
without initial r-enhancement.
\\
Allen {\em et al.\/} (\cite{allen12}) revised five stars already known in literature
and studied seven new objects (red symbols).
CS 29528--028 by Aoki {\em et al.\/} (\cite{aoki07}) is now classified as CEMP-s/r stars
owing to the high Eu newly detected by Allen {\em et al.\/} (\cite{allen12}; [Eu/Fe] = 2.16). 
CS 29503--010, for which only Ba was available among the s-elements (Aoki {\em et al.\/} 
\cite{aoki07}), also belongs to the CEMP-s/r stars, with [Eu/Fe] = 1.69. Discrepant 
abundances have been found by the two analyses ($\Delta$[Fe/H] = 0.3 dex), with a
strongly reduced metallicity: CS 29503--010 was classified as a CH disk star 
([Fe/H] = $-$1.09; Aoki {\em et al.\/} \cite{aoki07}), while Allen 
{\em et al.\/} (\cite{allen12}) find [Fe/H] = $-$1.69. A more recent study by Yong
{\em et al.\/} (\cite{yong13}) provide [Fe/H] = $-$1.39, with [Ba/Fe] = 1.51 more
consistent with Aoki {\em et al.\/} (\cite{aoki07}).
The CEMP-s/r main-sequence stars CS 29526--110 and CS 22183--015 were studied by Cohen {\em et al.\/} 
(\cite{cohen06}) and Aoki {\em et al.\/} (\cite{aoki02,aoki07}). For CS 29526--110,
Allen {\em et al.\/} (\cite{allen12}) determine s-and r-enhancements stronger than
Aoki {\em et al.\/} (\cite{aoki02}) ([ls/Fe] = 1.54, [hs/Fe] = 2.38 and [Eu/Fe] = 2.28, instead of 
[ls/Fe] = 1.00, [hs/Fe] = 1.88 and [Eu/Fe] = 1.73). The opposite result is seen for CS 22183--015:
Allen {\em et al.\/} (\cite{allen12}) considered this star a giant, as found by Johnson \& Bolte 
(\cite{JB02}), with [ls/Fe] = 0.64, [hs/Fe] = 1.53 and [Eu/Fe] = 1.37, while Cohen {\em et al.\/} 
(\cite{cohen06}) classified CS 22183--015 as a main-sequence star with
[ls/Fe] = 0.55, [hs/Fe] = 1.76 and [Eu/Fe] = 1.70.
In perfect agreement are instead the abundances of the CEMP-s/r CS 22898--027,
with [ls/Fe] = 0.9, [hs/Fe] = 2.2 and [Eu/Fe] 1.9 (Allen {\em et al.\/} \cite{allen12};
Aoki {\em et al.\/} \cite{aoki02}).
Other objects analyzed by Allen {\em et al.\/} (\cite{allen12}) are:
the CEMP-s giant CS 29512--073 ([Fe/H] = $-$2.06), the two CEMP-s BS 16077--077 and  
BS 16080--175, a CEMP-s/r CS 22887--048, and the CEMP+s/r giant BS 17436+058.
Two of them were members of the sample first presented by Tsangarides (\cite{tsangarides05}):
BS 17436+058, which is now classified as CEMP-s/r owing to the higher revised Eu
abundance ([Eu/Fe] = 1.13), and BS 16080--175.
\\
Placco {\em et al.\/} (\cite{placco13}) provide a detailed analysis of two newly discovered
stars, the CEMP-s HE 2138--3336 (black plus symbol) and the CEMP-s/r HE 2258--6358 ([Eu/Fe] 
= 1.68; violet diamond).
HE 2138--3336 ([Fe/H] = $-$2.79) has the highest [Pb/Fe] abundance ratio measured so far 
if non-local thermodynamic equilibrium corrections are included ([Pb/Fe] = +3.84), second only to
the CEMP-s/r CS 29497--030 ([Pb/Fe] = 3.65; Ivans {\em et al.\/} \cite{ivans05}).‬
\\ 
Cui {\em et al.\/} \cite{cui13} 
discovered a new CEMP-s/r giant, HE 1405-0822 with [Eu/Fe] = 1.54 (green diamod).
A large number of both s- and r- process elements are detected for this star, including
two lines for Nb ([Nb/Fe] = 0.98 $\pm$ 0.30). As Tc, Nb is an indicator of the binarity of the
stars. This is the second CEMP-s/r star with Nb determination (see CS 29497--030
by Ivans {\em et al.\/} \cite{ivans05}). [Nb/Zr] = 0.18 supports the binary scenario.
\\
Matrozis {\em et al.\/} (\cite{matrozis12}; brown diamond) provided an independent analysis 
for the CEMP-s/r giant HD 209621 by Goswami \& Aoki (\cite{goswami10}). 
While metallicity and effective temperature found 
by both works are in agreement ([Fe/H] = $-$1.9; $T_{\rm eff}$ = 4400 and 4500 K), the gravity 
estimated by Matrozis {\em et al.\/} \cite{matrozis12} is much lower
than that found by Goswami \& Aoki (\cite{goswami10}; log $g$ = 1.0 and 2.0 cgs). 
The discrepancy previously detected among the light s-elements ([Zr/Y] $\sim$ 1.4) has been 
substantially reduced ([Zr/Y] = 0.5), in better agreement with theoretical AGB models. 
\\
Pereira {\em et al.\/} (\cite{pereira12}) study the new high-velocity CH giant CD-62$^o$1346 
with [Fe/H] = $-$1.6 and high carbon and s-element abundances (brown 'x').
They also discuss HD 5223 (Goswami {\em et al.\/} \cite{goswami06}), showing that it is another 
example of a high-velocity CH star that exceeds the Galactic escape velocity.
Unfortunately, no Eu abundance is given for these stars.
\\
G 24--25 is a CH metal-poor subgiant ([Fe/H] = $-$1.4) studied by Liu {\em et al.\/} 
(\cite{liu12}; yellow diamod),
with a period of P = 3452 $\pm$ 67 days (Latham {\em et al.\/} \cite{latham02}).

It is noteworthy that SDSS J1707+58 is now discovered to be an RR Lyrae star 
(Kinman {\em et al.\/} \cite{kinman12}; see Stancliffe {\em et al.\/} \cite{stancliffe13} 
for a theoretical interpretation), instead of a main-sequence CEMP-s star (Aoki {\em et al.\/} 
\cite{aoki08}).
SDSS J1707+58 is one of the most metal-poor RR Lyrae stars known to date with [Fe/H] = $-$2.92,
about 0.4 dex lower than that found by Aoki {\em et al.\/} (\cite{aoki08}). It shows a high s-process abundance,
with [Ba/Fe] = 2.83 $\pm$ 0.51 (Sr has larger errors with [Sr/Fe] = 0.75 $\pm$ 0.65).
Owing to the limited number of revised observations, SDSS J1707+58 is not included in Fig.~\ref{Fig1}.
Another RR Lyrae star showing mild s-process enhancement is TY Gru (Preston {\em et al.\/} 
\cite{preston06}).

Additional newly discovered CEMP-s stars (for which only Sr and Ba among s-elements are detected)
are:
ten stars by Aoki {\em et al.\/} \cite{aoki13},
SDSS J0002+2928, SDSS J0126+0607, SDSS J1245-0738, SDSS J1349-0229, SDSS J1734+4316,
SDSS J1836+6317, 
SDSS J1626+1458, 
as well as SDSS J0711+6702, 
SDSS J1036+1212, 
SDSS J1646+2824, 
having enhanced Ba but subsolar [Sr/Fe].
Among them, three exhibit large excesses of Ba ([Ba/Fe] $>$ 1; SDSS J1836+6317 and SDSS J1734+4316;
SDSS J0126+0607 has Ba = +3.2).
One exhibits a moderate excess of Ba ([Ba/Fe] = +0.8; SDSS J0711+6702).
SDSS J1836+6317 and SDSS J1245−0738 are CEMP-s stars with large excesses of Na and Mg.
Two new CEMP-s have been analysed by Spite {\em et al.\/} (\cite{spite13}),
SDSS J111407.07+182831.7 and SDSS J114323.42+202058.0 (Fe/H $<$ -3), very likely 
belonging to binary systems.
Two additional possible CEMP-s are analyzed by Yong {\em et al.\/} (\cite{yong13}):
HE 0207−1423 and 52972-1213-507. 
Note that for these last stars no information is available among r-process elements.
 For accurate analyses additional data are needed.

The number of CEMP-s/r stars has increased from seventeen to twenty-four.
By looking at the stars having not undergone the first dredge-up episode (top panels),
the range covered by the averaged [hs/ls] ratios in CEMP-s/r stars (large symbols) 
seems to agree better with that of CEMP-s stars (small symbols) if the  
stars by Allen {\em et al.\/} (\cite{allen12}) are included in the analysis.
This may suggest that similar [hs/Fe] are shown by main-sequence CEMP-s and CEMP-s/r stars.
However, we strongly highlight the discrepant abundances obtained for some stars by
different authors (e.g., CS 29503--010 with $\Delta$[Fe/H] = 0.6 dex). 
As discussed by Allen {\em et al.\/} (\cite{allen12}), different authors employ
different procedures which could involve different line lists, atomic data for the lines 
used, adopted solar abundances, model atmospheres, and spectrum synthesis codes.
This suggests a sort of caution in 
combining data from the literature, and we propose that detailed analyses of individual stars 
are the more suitable starting point to improve theoretical AGB models and to understand the 
discrepancies still present between theory and observations (see Bisterzo {\em et al.\/} 
\cite{bisterzo12}).

An updated study of the CEMP-s and CEMP-s/r stars recently discovered is planned.
Promising spectroscopic results are expected in the near future (e.g., GAIA-ESO mission, 
as well as SkyMapper and LAMOST surveys, or the planned TMT facility).
This will significantly increase the number of newly discovered stars, which are observed
with high resolution, and will help to shed light on the presently debated issues
on the origin of CEMP-s and CEMP-s/r stars.

{\bf Acknowledgements:} 
S.B. wishes to acknowledge JINA for financial support under ND Fund
$\#$201387 and 305387.
Numerical calculations have been supported by B2FH Association (http://www.b2fh.org/).


\end{document}